    \documentclass[prd,aps,preprint,tightenlines,showpacs,nofootinbib,superscriptaddress]{revtex4-1}
\usepackage{mathrsfs}
\usepackage{amsfonts}
\usepackage{amsmath}
\usepackage{amssymb}
\usepackage{array}
\usepackage{verbatim}
\usepackage{bm}
\usepackage{epsfig}
\usepackage{graphicx,color}
\usepackage{relsize}
\usepackage{lineno}
\usepackage{float}
\usepackage{multirow}
\RequirePackage{xspace}

\def\lsim{\raise0.3ex\hbox{$<$\kern-0.75em\raise-1.1ex\hbox{$\sim$}}}

\def\gsim{\raise0.3ex\hbox{$>$\kern-0.75em\raise-1.1ex\hbox{$\sim$}}}

\newcommand{\be}{\begin{equation}}

\newcommand{\ee}{\end{equation}}

\def\beq{\begin{equation}}

\def\eeq{\end{equation}}

\def\beqa{\begin{eqnarray}}

\def\eeqa{\end{eqnarray}}

\newcommand{\ba}{\begin{eqnarray}}

\newcommand{\ea}{\end{eqnarray}}

\def\gappeq{\mathrel{\rlap {\raise.5ex\hbox{$>$}}

{\lower.5ex\hbox{$\sim$}}}}

\def\lappeq{\mathrel{\rlap{\raise.5ex\hbox{$<$}}

{\lower.5ex\hbox{$\sim$}}}}

\def\Toprel#1\over#2{\mathrel{\mathop{#2}\limits^{#1}}}

\begin{document}

\title{Neutrino trident scattering  at the LHC energy regime}

\vspace{3cm}

\author{Reinaldo {\sc Francener}}
\email{reinaldofrancener@gmail.com}
\affiliation{Instituto de Física Gleb Wataghin - UNICAMP, 13083-859, Campinas, SP, Brazil. }

\author{Victor P. {\sc Gon\c{c}alves}}
\email{barros@ufpel.edu.br}
\affiliation{Institute of Physics and Mathematics, Federal University of Pelotas, \\
  Postal Code 354,  96010-900, Pelotas, RS, Brazil}
\affiliation{Institute of Modern Physics, Chinese Academy of Sciences,
  Lanzhou 730000, China}

\author{Diego R. {\sc Gratieri}}
\email{drgratieri@id.uff.br}
\affiliation{Escola de Engenharia Industrial Metal\'urgica de Volta Redonda,
Universidade Federal Fluminense (UFF),\\
 CEP 27255-125, Volta Redonda, RJ, Brazil}
\affiliation{Instituto de Física Gleb Wataghin - UNICAMP, 13083-859, Campinas, SP, Brazil. }

\vspace{3cm}

\begin{abstract}
The neutrino trident  scattering process in  neutrino - tungsten interactions at the LHC energy regime is investigated, and the cross-sections for different leptonic final states in coherent and incoherent interactions are estimated. Furthermore, the associated number of events at FASER$\nu$2 detector is estimated considering different predictions for the flux of incident neutrinos on the detector, based on distinct hadronic models for the particle production in $pp$ collisions at ultra-forward rapidities.  Our results indicate that the observation of the neutrino trident process is, in principle, feasible  at the Forward Physics Facility.
\end{abstract}


\keywords{Neutrino tridents; TeV-energy neutrinos; Forward Physics Facility.}

\maketitle

\vspace{1cm}


The recent detection of  neutrinos 
by the SND@LHC \cite{SNDLHC:2023pun} and FASER$\nu$ \cite{FASER:2021mtu,FASER:2023zcr} experiments has started the far - forward neutrino physics program at the Large Hadron Collider (LHC), which 
is expected to be boosted during the high luminosity LHC (HL - LHC) era with the experiments to be installed in the 
Forward Physics Facility (FPF) \cite{Anchordoqui:2021ghd,Feng:2022inv}.
The basic idea is that a large number of neutrinos originate through the decay of hadrons produced at forward rapidities in high energy $pp$ collisions and, consequently, LHC provides an intense and strongly collimated  beam of high - energy neutrinos  that can be used to study neutrino - hadron interactions in a detector centered around the beam - collision axis, as e.g. FASER$\nu$, FASER$\nu2$ and FLArE, or off axis as e.g. SND@LHC.   In recent years, several studies have explored the possibility of
probing Standard Model (SM) predictions,  constrain the neutrino - nucleon cross-sections in an explored energy range and searching for signals of New Physics using the forthcoming results from neutrino - hadron interactions at the FPF (For a review see, e.g., Ref. \cite{Anchordoqui:2021ghd}), strongly motivated by the huge integrated luminosity predicted for the HL - LHC run of ${\cal{L}} = 3$ ab$^{-1}$. Such luminosity is expected to allow us to investigate rare processes, characterized by tiny cross-sections, whose predictions have not yet tested experimentally, or that current data is still scarce. 

\begin{figure}[t]
	\centering
	\begin{tabular}{ccccc}
\includegraphics[width=0.48\textwidth]{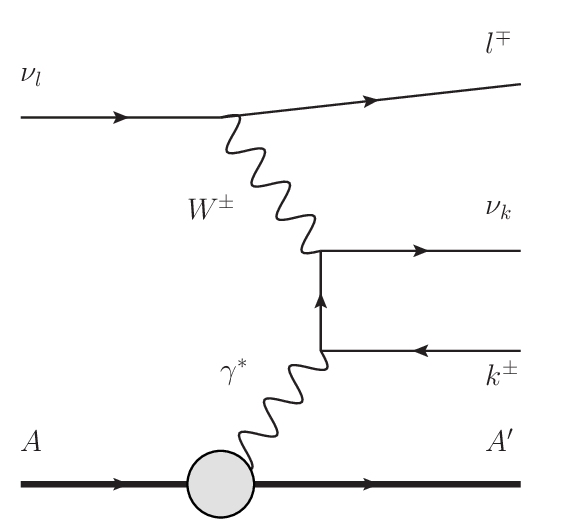} &
\includegraphics[width=0.48\textwidth]{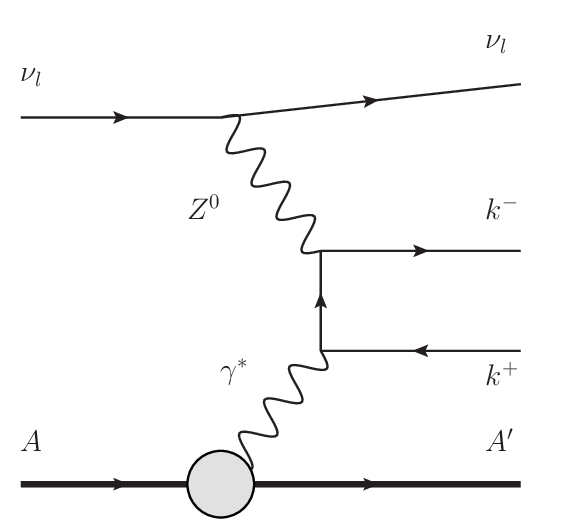} 
	\end{tabular}
\caption{ Feynman diagrams for the neutrino trident scattering off a nucleus target associated with a $W^\pm$ (left) and $Z^0$ (right) exchange. The lepton flavors $l$ and $k$ can be equal or different. For a $W^\pm$ exchange, the charged lepton pair can be constituted by different flavors.  }
\label{fig:diagrams}
\end{figure}

In this letter, we will investigate the neutrino trident production, which is a weak process characterized by the production of a pair of charged leptons through the neutrino scattering in the Coulombian field of a heavy nucleus.  Such process can be mediated by a $W^{\pm}$ or a $Z^0$ boson exchange, as represented in Fig. \ref{fig:diagrams}, and as the mass of these gauge boson is much larger than the relevant momentum transfer, the cross-section can be accurately estimated assuming a four lepton contact interaction. For a pair of muons in the final state, the trident process $\nu_{\mu} \rightarrow \nu_{\mu} \mu^+ \mu^-$  has been observed by CHARM-II \cite{CHARM-II:1990dvf}, CCFR \cite{CCFR:1991lpl} and NuTeV \cite{NuTeV:1999wlw} experiments, with the results in agreement with the SM predictions. However, measurements for other incident neutrinos and/or final states have not yet obtained.  In recent years, several theoretical studies  have been carried out, showing the feasibility of observing the trident process in experiments such as DUNE \cite{Altmannshofer:2019zhy}, IceCube \cite{beacom2,Sarkar:2023dvr} and  near detectors \cite{Ballett:2018uuc}. Our goal in this study is to provide, for the first time, the predictions for the trident cross-sections at the LHC energies and the event rates at the FASER$\nu2$ detector for a variety of neutrino and antineutrino - induced trident processes. As we will demonstrate below, our results indicate that a future experimental analysis of neutrino trident process at the FPF is, in principle, feasible.

In what follows, we will present our results for trident production induced by electronic and muonic (anti)neutrinos, considering the neutrino - tungsten interaction  in the kinematical regime  that will be covered by FASER$\nu$2 detector, which is a detector based on the emulsion technology, with a transverse size of 40 cm $\times$ 40 cm and total tungsten mass of 20 tons, centered around the beam - collision LHC axis. In our analysis, we will estimate the cross-section using the four lepton contact interaction theory, as described e.g. in Refs. \cite{Ballett:2018uuc,Altmannshofer:2019zhy}, and taken into account of contribution of coherent and incoherent elastic scatterings, where the leptonic system scatters on the full nucleus or with the individual nucleons inside the nucleus, respectively. In particular, we will present separately the incoherent predictions for the interaction with a proton and with a neutron. It is important to emphasize that these two contributions are characterized by distinct final states: while in a coherent scattering the nucleus remains intact, in the incoherent one, it is expected to breakup, generating a hadronic activity in addition to the leptonic system, which is  associated with the fragments of the nucleus. Such distinct topology can be used, in principle, to separate the coherent and incoherent events. As in Ref. \cite{Altmannshofer:2019zhy}, the results will be calculated considering the full $2 \rightarrow 4$ scattering process, without assume the validity of the equivalent photon approximation.

\begin{figure}[t]
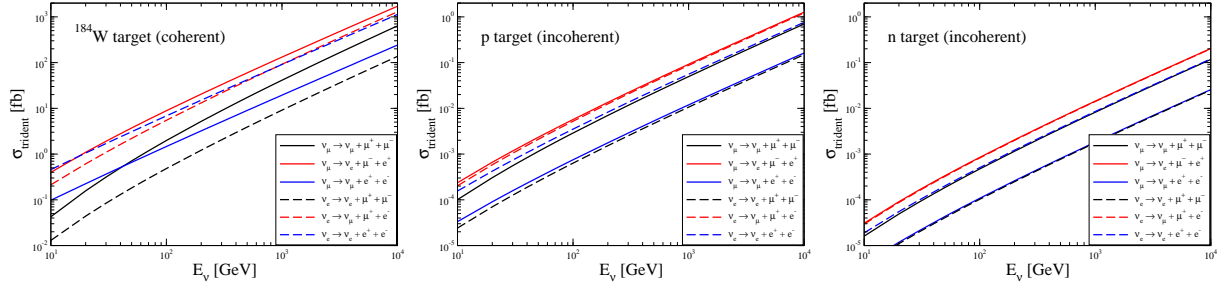

	\centering
	\begin{tabular}{ccccc}
\includegraphics[width=0.32\textwidth]{sigma_nu_W.eps} &
\includegraphics[width=0.32\textwidth]{sigma_nu_p.eps} &
\includegraphics[width=0.32\textwidth]{sigma_nu_n.eps} 
	\end{tabular}
\caption{Cross-sections for the trident processes as a function of the incident neutrino energy. We consider interactions of muonic (solid) and electronic (dashed) neutrinos originating muon pair (black) muon plus electron (red) and electron pair (blue). Our results are for coherent scattering with the full nucleus (left panel) and for incoherent scattering with a proton (middle panel) and a neutron (right panel) inside the nucleus. }
\label{fig:cs}
\end{figure}

Initially,  in Fig. \ref{fig:cs}, we present the cross-sections for neutrino-induced trident processes, derived using the formalism described in detail in Refs. \cite{Ballett:2018uuc,Altmannshofer:2019zhy} that was implemented in the Monte Carlo event generator proposed in  Ref. \cite{Altmannshofer:2019zhy} to estimate the trident events at DUNE. In our analysis, we have adapted this event generator for a tungsten target. We will present predictions for $\sigma_{\nu_l A}$, but we also have estimated the cross-sections for an incident antineutrino and obtained similar results. 
The predictions in Fig. \ref{fig:cs} are for the coherent scattering (left) and for the incoherent interaction with a proton (middle) and with a neutron (right).
 Black, red, and blue lines represent final states with two muons, muon plus electron, and two electrons, respectively. One has that the $\nu_{\mu} \rightarrow \nu_{e} \mu^- e^+$ process has the largest cross-section over the energy range considered. In contrast, the $\nu_{e} \rightarrow \nu_{e} \mu^+ \mu^-$ process has the smallest cross-section. It is important to emphasize that the cross-section of the  $\nu_{\mu} \rightarrow \nu_{\mu} \mu^+ \mu^-$ process receives the contribution of both diagrams presented in Fig. \ref{fig:diagrams}, which implies a destructive interference that reduces the magnitude of the cross-section. 
  Our results also indicate that
 the coherent scattering dominates, which is directly associated with the $Z^2$ factor present in the cross-section. Moreover, one has that the cross-section for the incoherent scattering with a proton is one order of magnitude larger than for a neutron target, which is smaller since it is electrically neutral. These conclusions are in agreement with those presented in Ref. \cite{Altmannshofer:2019zhy}  for smaller neutrino energies.

\begin{figure}[t]
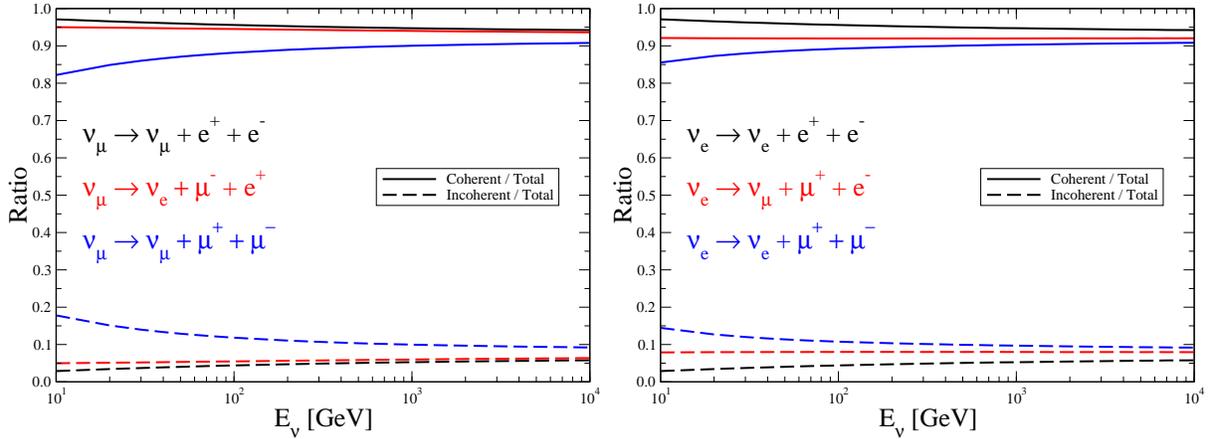

	\centering
	\begin{tabular}{ccccc}
\includegraphics[width=0.48\textwidth]{ratio_num.eps} &
\includegraphics[width=0.48\textwidth]{ratio_nue.eps} 
	\end{tabular}
\caption{ Ratio between coherent (solid) and incoherent (dashed) contributions to the total cross-section for trident events induced by muonic (left) and electronic (right) neutrino, considering  a tungsten target. }
\label{fig:ratio}
\end{figure}

In order to quantify the contribution of coherent and incoherent scatterings for a given final state, in Fig. \ref{fig:ratio} we show the energy dependence of the ratio between these contributions to the total cross-section, which is the sum of both processes, considering  $\nu_\mu \,^{187}W$ (left) and $\nu_{e} \,^{187}W$ interactions. 
In agreement with the results presented in Fig. \ref{fig:cs},  the coherent process dominates for all  channels in the energy range considered. The incoherent contribution is more significant for the production of a pair of muons than for the production of a $e^+e^-$ system, which is associated with the fact that a larger value of the photon virtuality is needed in order to produce a muon pair. In this case, the probability of probing the internal structure of the nucleus is larger, which implies a larger cross-section for the incoherent interaction. It also explains  why the production of muon plus electron has an intermediate contribution from the incoherent case compared to the cases mentioned before. Our results indicate that the contribution of incoherent processes is smaller than 10\% in the LHC energy range.

The cross-section is one of the main ingredients for estimating the number of trident of events in a given detector. Another important ingredient is the flux of incident neutrinos. In our analysis, we will consider  the neutrino flux incident on the FASER$\nu$2 estimated  in Ref.\cite{Kling:2023tgr} considering distinct modeling for the hadron production at large pseudorapidities in $pp$ collisions at the LHC energy, implemented in different Monte Carlo generators. As demonstrated in Ref.\cite{Kling:2023tgr}, the main contributions for the neutrino fluxes comes from the  decay of light mesons (pions and kaons) and from  the decay of charmed hadrons.
In order to quantify the uncertainty in our predictions, we will estimate the neutrino flux associated with the decay of light mesons using the following MC  generators: \texttt{EPOS-LHC} \cite{Pierog:2013ria}, \texttt{DPMJET 3.2019.1} \cite{Roesler:2000he,Fedynitch:2015}, \texttt{QGSJET II-04} \cite{Ostapchenko:2010vb}, \texttt{SIBYLL 2.3d} \cite{Ahn:2009wx,Riehn:2015oba} and \texttt{Pythia 8.2} \cite{Fieg:2023kld,Sjostrand:2014zea}. On the other hand,  the neutrino flux coming from charmed hadrons will be estimated using the results derived in Refs.   \cite{Bai:2020ukz,Bai:2021ira,Bai:2022xad} (denoted \texttt{BDGJKR}),  \cite{Buonocore:2023kna} (denoted \texttt{BKRS}),  \cite{Bhattacharya:2023zei} (denoted \texttt{BKSS} $k_T$), \cite{Maciula:2022lzk}  (denoted \texttt{MS} $k_T$)  and  \cite{Ahn:2011wt,Fedynitch:2018cbl} (denoted \texttt{SIBYLL 2.3d}). In particular, we will use the same generator combinations  for the total neutrino flux considered in Ref. \cite{Kling:2023tgr}.

\begin{figure}[t]
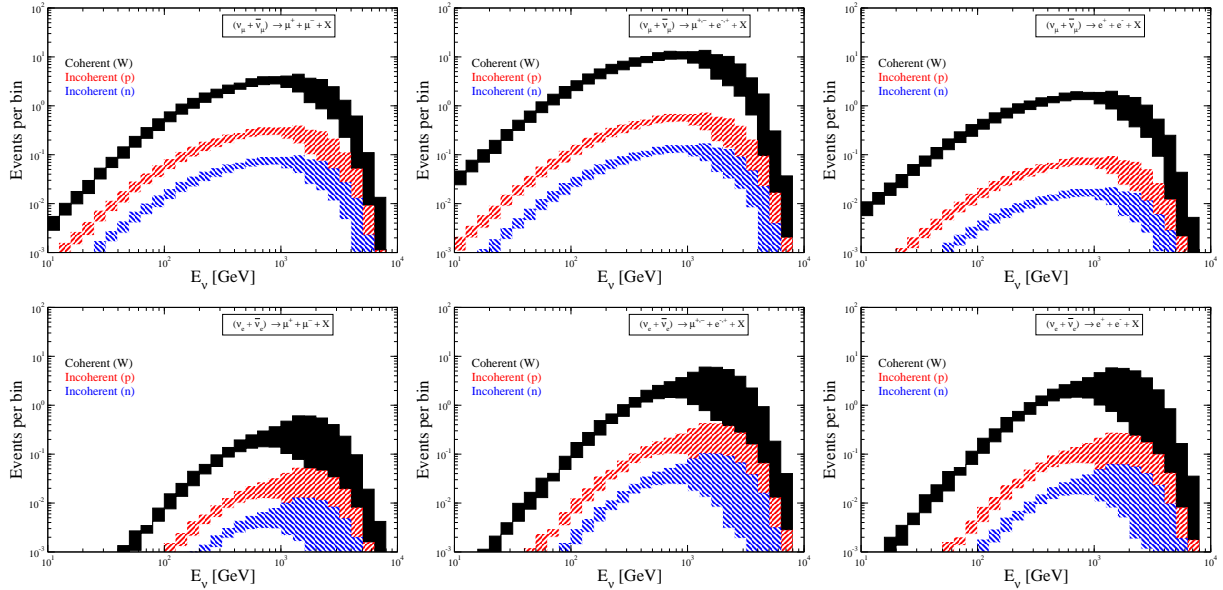

	\centering
	\begin{tabular}{ccccc}
\includegraphics[width=0.32\textwidth]{trident_num_mu+_mu-.eps} &
\includegraphics[width=0.32\textwidth]{trident_num_mu+-_e-+.eps} &
\includegraphics[width=0.32\textwidth]{trident_num_e+_e-.eps} \\
\includegraphics[width=0.32\textwidth]{trident_nue_mu+_mu-.eps} &
\includegraphics[width=0.32\textwidth]{trident_nue_mu+-_e-+.eps} &
\includegraphics[width=0.32\textwidth]{trident_nue_e+_e-.eps} 
	\end{tabular}
\caption{ Number of trident events per bin expected in the FASER$\nu$2 detector,  induced by muonic (top) and electronic (bottom) neutrinos, derived assuming the expected luminosity for the HL-LHC of $3\, \mathrm{ab}^{-1}$. Our results are for different leptons in the final state: muon pair (left), muon plus electron (middle) and electron pair (right). The uncertainty band was constructed considering different Monte Carlo generators for the incident neutrino flux. }
\label{fig:events1}
\end{figure}

The number of neutrino-induced events at the FASER$\nu$2 detector   is given by the product of the time-integrated neutrino flux with the neutrino interaction probability, which is defined  by $\sigma_{\nu A} \rho L/m_{A}$, where $\sigma _{\nu A}$ is the neutrino-nucleus cross-section, $ \rho$ the density of the target material ($19.3\,\mathrm{g/cm}^{3}$ for tungsten), $L$ the length of the detector ($6.6$ m) and $m_{A}$ the mass of the target nucleus.
In Fig. \ref{fig:events1} we present our results for the number of trident events per bin expected in the  FASER$\nu$2 detector during the high luminosity LHC  regime for events induced by $\nu_\mu$ (top) and $\nu_e$ (bottom). We consider three leptonic final states: muon pair (left), muon plus electron (middle), and electron pair (right). Furthermore, we present separately the coherent and incoherent scattering contributions. The uncertainty band was constructed with the different MC generators mentioned previously, limiting the maximum and minimum values with the largest and smallest predictions for the neutrino flux.
We predict a larger number of trident events induced by muonic neutrinos, with the production of charged leptons of different flavors being dominant. One has that the uncertainty bands increase appreciably for events with $E_{\nu}  \gtrsim 10^3\, \mathrm{GeV}$. Moreover, the uncertainty at high energies is larger for  trident processes induced by electronic neutrinos, since it receive a  greater contribution from the decay of charmed hadrons, whose predictions from the different production models still are largely distinct (See, e.g. Ref. \cite{Kling:2023tgr}). As expected from the analysis of the cross-sections, our results also indicate that coherent interactions dominate the number of events, about an order of magnitude above the incoherent contribution.

Finally, in Fig. 5 we present our results for  the expected number of trident events in the FASER$\nu$2 detector characterized by a given final state, independently of the flavor of the incident (anti)neutrino. We have summed the coherent and incoherent cross-sections, as well as the contributions associated with the neutrino and antineutrino-tungsten interactions. Our results are presented 
for a muon pair (left panel), a muon plus electron (middle panel) and an electron pair (right panel) considering different MC generators for the incident (anti)neutrino flux. One has that the number of events per bin for $E_\nu \approx 10^3\, \mathrm{GeV}$ is larger than $3$ for a pair of leptons of the same flavor, and  larger than $10$ for a muon plus electron final state. Such results indicate that the observation of the trident process is, in principle, feasible  in the FASER$\nu$2 detector.

\begin{figure}[t]
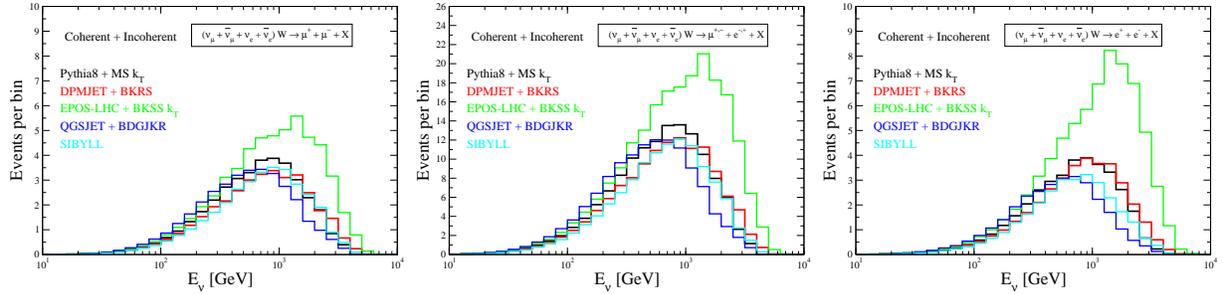

	\centering
	\begin{tabular}{ccccc}
\includegraphics[width=0.32\textwidth]{trident_total_mu+_mu-.eps} &
\includegraphics[width=0.32\textwidth]{trident_total_mu+-_e-+.eps} &
\includegraphics[width=0.32\textwidth]{trident_total_e+_e-.eps} 
	\end{tabular}
\caption{Total number of trident events in the  FASER$\nu$2 detector characterized by a given final state, estimated considering  different MC generators for the incident neutrino flux. Results for a muon pair (left panel), a muon plus electron (middle panel) and an electron pair (right panel). }
\label{fig:events2}
\end{figure}


As a summary, in this letter we have performed an exploratory study of the trident process in LHC energy regime. We estimate the total cross-sections  considering the production of different lepton final states. Moreover, we have calculated the expected number of events for the  high luminosity LHC regime, using different Monte Carlo generators to estimate the incident neutrino flux. We have demonstrated that, in principle, this rare process can be observed in the FASER$\nu$2 detector, which will allow us to test the SM predictions as well as searching for the contribution of new gauge bosons predicted in some scenarios for beyond the standard model physics. The results presented here strongly motivate a more detailed analysis in terms of angular and energetic distributions of the charged leptons present in the final state, as well as the analysis of potential backgrounds. We plan to  perform these studies in  future publications.

\begin{acknowledgments}

R. F. acknowledges support from the Conselho Nacional de Desenvolvimento Cient\'{\i}fico e Tecnol\'ogico (CNPq, Brazil), Grant No. 161770/2022-3. V.P.G. was partially supported by CNPq, FAPERGS and INCT-FNA (Process No. 464898/2014-5). D.R.G. was partially supported by CNPq and MCTI.

\end{acknowledgments}

\hspace{1.0cm}

\end{document}